\documentclass[prb]{revtex4-1} 

\usepackage{amsmath}  
\usepackage{amsfonts} 
\usepackage{graphicx} 
\usepackage{esint} 
\usepackage{color} 
\usepackage{soul} 

\begin{document}


\title{The Electric Force Between Two Straight Parallel Resistive Wires Carrying DC-Currents {in the Asymptotic Limit of Infinitely Thin Wires} }

\author{Udo Ausserlechner \\ https://orcid.org/0000-0002-8229-9143}
\email{udo.ausserlechner@infineon.com} 
\affiliation{Infineon Technologies, Sense and Control, Siemensstrasse 2, 9500 Villach, Austria}



\date{\today}

\begin{abstract}
During the years 1948-2019 the {ampere} was defined via the \emph{magnetic force} between two long thin parallel wires carrying stationary current. However, if a stationary current flows through a \emph{resistive} wire, static electric charges appear on the surface of the wire, and this will lead to an additional \emph{electric force} between the wires. This article discusses the ratio of electric over magnetic forces in the asymptotic limit of infinitely thin wires, which is not accessible by numerical methods. 
The electric force between the two wires depends also on the choice of the common ground node. \emph{For extremely thin or extremely long resistive wires the electric force dominates over the magnetic one.}
\end{abstract}

\maketitle 

\section{Introduction} 

In the International System of Units (SI) electric current is measured in the unit of ampere (A). Until recently (2019) the ampere was defined via the \emph{magnetic} force between two current carrying wires \cite{BeckerSauter,JacksonBuch,Hallen,Simonyi,Hofmann,Lehner}:
\begin{equation}\label{old-amp-definition}\begin{split}
& \text{The ampere is that constant current which, if maintained in two straight parallel} \\ 
& \text{conductors of infinite length, of negligible circular cross-section, and placed one} \\ 
& \text{metre apart in vacuum, would produce between these conductors a force equal} \\ 
& \text{to } 2\times 10^{-7} \text{ newtons per metre of length.} 
\end{split}\end{equation}
Reading this definition one naturally assumes that we only have to make the wire long and thin enough to get a current measurement with arbitrarily high precision. However, for resistive wires this is utterly wrong! \emph{In this article I will show that the longer and the thinner the wire becomes the more this definition of the ampere is corrupted by an additional electric force between the wires.} 
{Yet, this paper does not intend to question the accuracy of state-of-the-art current measurements, which use different and more sophisticated geometries than in} (\ref{old-amp-definition}), see Refs. \onlinecite{Nakamura1978,RobinsonSchlamminger2016,SchlammingerHaddad2019}. Conversely, I intend to show that, without any additional counter-measures against electric forces (such as two electrical shields), the simple geometry of two straight parallel wires like in (\ref{old-amp-definition}) gives \emph{infinite error} in the asymptotic limit of infinitely thin resistive wires. This was already shown in Ref. \onlinecite{Assis2wire}, however, there the authors neglected the infinite charges at the ends of the wires, whereas this article gives a rigorous {theory for wires of finite length} in the limit of vanishing diameter. 
I do not know why electric forces were not mentioned in (\ref{old-amp-definition}), although it is well known that a DC current through a conductor generates an electric field inside \emph{and outside} the conductor \cite{Marcus1941,Stratton1941,Merzbacher1980,Jackson1996,Assis1999,AssisHernandes2007,Zangwill}. One reason might be a lack in awareness of surface charges in electric circuits.

\section{Current Flow Through A Prolate Ellipsoid}
\label{sec:ellipsoid}

\begin{figure}[t]
\centering
\includegraphics[width=0.40\textwidth]{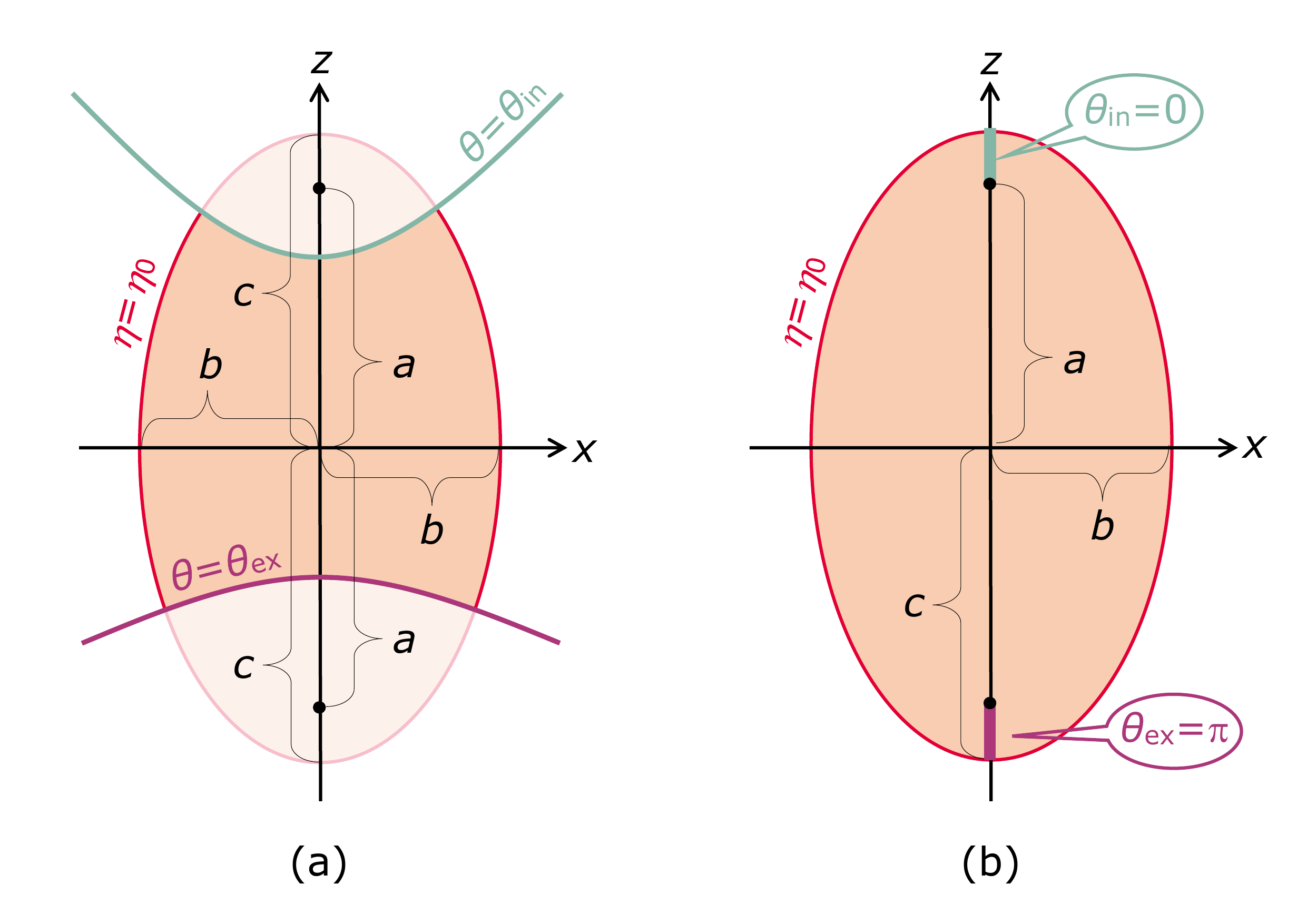}
\caption{(a) Cross-section of a prolate ellipsoid with its surface in $\eta=\eta_0$. The $z$-axis is the axis of rotational symmetry. The contact of current input is in $\theta=\theta_\mathrm{in}$, and the contact of current output is in $\theta=\theta_\mathrm{ex}$. The current flows through the darker orange region. The focus points are on the $z$-axis in $z=\pm a$. The vertex points are on the $z$-axis in $z=\pm c$. The long axis has length $2c$ and the maximum diameter is $2b$, with $c^2=a^2+b^2$. (b) shows the ellipsoid with line contacts in $\theta=0$ (for current input) and in $\theta=\pi$ (for current output). The contacts reach from each focus point to the closer vertex point. }
\label{fig:ellipsoid3}
\end{figure}

{In this paper I model the wires as prolate ellipsoids with DC-currents flowing along their long axes. Although ellipsoidal coordinates have been used to study electrostatic charging of thin wires} \cite{Maxwell,Andrews,Jackson-revisited}, {no one seems to have used them to study stationary current flow in thin wires of finite length. The ellipsoidal model allows for a rigorous mathematical treatment, and it accurately models cylindrical wires in the asymptotic limit of infinite slimness.} 

Figure \ref{fig:ellipsoid3}(a) shows a conductor in the shape of a prolate ellipsoid with the long axis of length $2c$ along the $z$-axis. The conductor volume is parametrized by $0\le\eta\le\eta_0$ with arbitrary $\theta$ and $\psi$ in an ellipsoidal reference frame $(\eta,\theta,\psi)$ defined by   
\begin{equation}\label{eq:1}\begin{split}
x & = a \sinh(\eta) \sin(\theta) \cos(\psi) , \quad 
y  = a \sinh(\eta) \sin(\theta) \sin(\psi) , \quad 
z  = a \cosh(\eta) \cos(\theta) , 
\end{split}\end{equation}
$(x,y,z)$ being the Cartesian coordinates. The ellipsoid has rotational symmetry---its cross-sections orthogonal to the $z$-axis are circles. The largest circle is in $z=0$, which is equivalent to $\theta=\pi/2$, with the diameter $2b:=2a\sinh(\eta_0)$. The length of the ellipsoid is $2c:=2a\cosh(\eta_0)$. Thus, the aspect ratio (slimness) of the ellipsoid is $c/b=\coth(\eta_0)$. It becomes infinite for $\eta_0\to 0$, $c/b\to\infty$. Then the ellipsoid degenerates to a straight line of length $2a$ with its ends being the foci of the ellipsoid in $(x,y,z)=(0,0,\pm a)$. I call this the thin wire limit.

\subsection{The Potential Inside the Ellipsoid and the Charge on the Contacts}
\label{sec:inside-ellipsoid}

Current flows between the hyperboloid coordinate surfaces $\theta=\theta_\mathrm{in}$ and $\theta=\theta_\mathrm{ex}$ for $\eta<\eta_0$ and arbitrary $\psi$. There, the potential is constant (Dirichlet boundary condition). The lateral surface of the truncated ellipsoid $\eta=\eta_0$, $\theta_\mathrm{in}\le\theta\le\theta_\mathrm{ex}$, arbitrary $\psi$, is insulating. There, the normal derivative of the potential vanishes (Neumann boundary condition). All boundary conditions do not depend on $\eta$ and $\psi$, therefore the same applies to the potential, $\phi_\mathrm{inside}(\theta)$. For stationary current flow with uniform scalar conductivity $\kappa$ the potential satisfies Laplace's differential equation in ellipsoidal coordinates with the solution \cite{MoonSpencer}
\begin{equation}\label{eq:6}
\phi_\mathrm{inside}(\theta) = c_1 Q_0(\cos(\theta)) + c_2 . 
\end{equation}
$Q_0(\cos(\theta))=\ln(\cot(\theta/2))$ is a Legendre {function} of the second kind and of order zero. 
We define ground potential in $\theta=\pi/2$, which means $
c_2 = 0 .
$
The constant $c_1$ is determined by the current $I_0$ flowing through the ellipsoid (Appendix \ref{sec:appL}).

For the potential close to the $z$-axis we write $r=\mathrm{d}r$ with $|\mathrm{d}r|<<b$. From (\ref{eq:1}) we know $r=a\sinh(\eta)\sin(\theta)$. Therefore small $r$ means either $\eta\to 0$ or $\theta\to 0$ or $\theta\to\pi$. 
Let us consider the case of 'small $\theta$'. There we have the full ellipsoid of Figure \ref{fig:ellipsoid3}(b), and the contacts are the lines $a\le |z|\le c$ in the shape of infinitely thin needles of infinite conductivity. They are on the $z$-axis, pushed into the ellipsoid until the tips of the needles reach the focus points. 
We enter $\theta=\mathrm{d}\theta$ with $|\mathrm{d}\theta|<<1$ into (\ref{eq:1}),
\begin{equation}\label{eq:24}\begin{split}
\mathrm{d}r & = a \sinh(\eta) \mathrm{d}\theta, \quad z  = a \cosh(\eta) \quad \Rightarrow \mathrm{d}r = \sqrt{z^2-a^2}\,\mathrm{d}\theta \quad \textrm{with } a\le z\le c. 
\end{split}\end{equation}
Inserting this into (\ref{eq:6}) gives for $a\le z\le c$ 
\begin{equation}\label{eq:25}\begin{split}
\phi_\mathrm{inside}(\mathrm{d}r, z) & = \frac{I_0}{4\pi\kappa (c-a)} \ln\!\left(4\,\frac{z^2-a^2}{(\mathrm{d}r)^2}\right)+c_2  + \mathcal{O}\left(\mathrm{d}r\right), \\  
E_z(r=0, z) & =  \frac{-I_0 z}{2\pi\kappa (c-a)}\;\frac{1}{z^2-a^2}, \\ 
E_r(\mathrm{d}r, z) & = \frac{I_0}{2\pi\kappa (c-a)}\;\frac{1}{\mathrm{d}r} + \mathcal{O}\left(\mathrm{d}r\right) .
\end{split}\end{equation}
Applying Gauss's electric flux theorem \cite{Gauss-theorem} to $E_r$ on the contact gives 
the charge on the contact
\begin{equation}\label{eq:27b}
q_\mathrm{cont}=\epsilon_0 I_0/\kappa . 
\end{equation}
With $\epsilon_0=8.854\times 10^{-12}\;\mathrm{Vs/Am}$, $I_0=1.06\;\mathrm{A}$, $\kappa=5.88\times 10^{7}\;\mathrm{S/m}$ (conductivity of copper), and the elementary charge $e=1.6\times 10^{-19}\;\mathrm{As}$ we find that there is only a \emph{single} electron on the contact, regardless if the ellipsoid is very slim or not! 
{Interestingly, the same tiny amount of surface charge} $q_\mathrm{cont}$ sits in a $90$° bend in an ordinary wire with conductivity $\kappa$ to guide the current $I_0$ around the corner \cite{Zangwill}. Hence, a very small charge has a huge effect on the local electric field. 

Irrespective of the slimness of the ellipsoid the potential at the contact of current injection in (\ref{eq:25}) is \emph{infinite}, because the contact has zero diameter (it is a line). Contacts with infinite contact resistance are not unusual in electrical engineering---they occur for example in Van der Pauw's measurement of the sheet resistance of thin plates with point or line contacts \cite{originalVdP} and they also occur in Hall plates with point or line contacts \cite{Hall}. If current is injected in such a contact, its potential rises unboundedly with a logarithmic singularity. A more surprizing finding is that in (\ref{eq:25}) the electric field $E_z$ on the contact and parallel to the contact does \emph{not} vanish. This is a consequence of the fact that the potential at the contact is infinite, which means that it takes infinite energy to shuffle a small test charge from infinity onto the contact. Conversely, the electric field along the contact is only finite, and therefore we need only finite energy to shift this test charge along the contact.
The finite energy needed to shift a test charge along the contact is still negligible against the infinite energy to bring it onto the contact, and in this asymptotic sense the contact is still an iso-potential surface. The radial field $E_r$ on the entire contact is infinite, see (\ref{eq:25}), and therefore the field lines of $\mathbf{E}$ are perfectly perpendicular to the contact---in this respect the contact still behaves as we expect it from an ideal contact. Note that on the one hand the conductivity of the line contact is infinite, on the other hand its thickness is zero, and this may serve as an explanation for the line contact not being able to force zero tangential electric field. Note also that we cannot study this phenomenon by numerical simulation techniques like finite elements (FEM), because these programs cannot handle infinite quantities. They cannot represent infinite radial field $E_r$ on the line contact and therefore erroneously they assume \emph{zero} tangential field $E_z=0$ to arrive at correct E-field lines orthogonal to the contacts.  
\newline The same phenomenon occurs in electrostatics, if we charge up an infinitely thin straight wire of finite length \cite{Jackson-revisited}: The capacitance of such an ideal needle vanishes, see (\ref{eq:superpos3}). Hence, it takes infinite energy to charge it. The charges distribute homogeneously on the ideal needle, yet they seem not to be in equilibrium, because there is a finite electric field acting on them in the direction of the needle. Again here the finite energy needed to shift a test charge along the ideal needle is negligible against the infinite energy needed to bring it onto the ideal needle, and the field lines are perfectly orthogonal to the ideal needle. Also here, standard FEM codes assume isopotential along the needle, thereby failing to predict the finite tangential electric field on the needle.

\subsection{The Potential Outside the Ellipsoid and the Charge on the Ellipsoid}
\label{sec:outside-ellipsoid}

The general ansatz for the potential outside the full ellipsoid of Fig. \ref{fig:ellipsoid3}(b) is
\begin{equation}\label{eq:40}\begin{split}
\phi_\mathrm{outside}(\eta,\theta) = \sum_{n=0}^\infty & \left( A_n P_n(\cos(\theta)) + B_n Q_n(\cos(\theta)) \right) Q_n(\cosh(\eta)) .
\end{split}\end{equation}
In (\ref{eq:40}) we have to discard the terms $Q_n(\cos(\theta))$, because they are singular in $\theta=0$ and $\theta=\pi$. This means $B_n=0$. Indeed, the potential is singular in $\theta=0$ and $\theta=\pi$, but only for $\eta=\eta_0$, whereas it is regular for $\eta>\eta_0$. 
Continuity of the potential at the surface $\eta=\eta_0$ means $\phi_\mathrm{inside}(\theta)=\phi_\mathrm{outside}(\eta_0,\theta)$. Splitting up this identity into even and odd functions of $\theta$ gives 
\begin{equation}\label{eq:41b}\begin{split}
0 & = \sum_{n=0}^\infty A_{2n} Q_{2n}(\cosh(\eta_0)) P_{2n}(\cos(\theta)) , \\ 
c_1 Q_0(\cos(\theta)) & = \sum_{n=0}^\infty A_{2n+1} Q_{2n+1}(\cosh(\eta_0)) P_{2n+1}(\cos(\theta)) .
\end{split}\end{equation}
Thus, $A_{2n}=0$. For $A_{2n+1}$ we multiply both sides with $P_{2m+1}(\cos(\theta)) \sin(\theta)$  and integrate over $\theta: 0\to\pi$. Note that $P_m(x), P_n(x)$ are orthogonal \cite{Arfken-ortho}, $\int_{-1}^1 P_m(x) P_n(x)\,\mathrm{d}x= 2\delta_{m,n}/(2n+1)$, but $P_m(x), Q_n(x)$ are \emph{not} orthogonal. In Appendix \ref{sec:appA} we prove 
\begin{equation}\label{eq:42b}
\int_0^\pi Q_0(\cos(\theta)) P_{2n+1}(\cos(\theta))\sin(\theta)\,\mathrm{d}\theta = \int_{-1}^1 Q_0(x) P_{2n+1}(x)\,\mathrm{d}x = \frac{1}{(n+1)(2n+1)}.
\end{equation}
Using this in (\ref{eq:41b}) finally gives 
\begin{equation}\label{eq:42}
A_{2n+1} = c_1\,\frac{4n+3}{2(n+1)(2n+1) Q_{2n+1}(\cosh(\eta_0))} > 0 .
\end{equation}

The electric field component perpendicular to the surface of the conductor is \cite{MoonSpencer3}
\begin{equation}\label{eq:45}\begin{split}
E_{\perp,\mathrm{outside}} & = \mathbf{E}_\mathrm{outside}\cdot\mathbf{n}_\eta = \frac{-1}{a\sqrt{(\sinh(\eta_0))^2+(\sin(\theta))^2}}\, \left.\frac{\mathrm{d}\phi_\mathrm{outside}}{\mathrm{d}\eta}\right|_{\eta_0} \\ 
& = \frac{-1}{a\sqrt{(\sinh(\eta_0))^2+(\sin(\theta))^2}}\, \sum_{n=0}^\infty A_n P_n(\cos(\theta)) \frac{\mathrm{d}}{\mathrm{d}\eta_0} Q_n(\cosh(\eta_0)) ,
\end{split}\end{equation}
where $\mathbf{n}_\eta$ is the unit vector in the direction of growing $\eta$ (with $\theta$ and $\psi$ staying constant). In (\ref{eq:45}) we can further use 
\begin{equation}\label{eq:46}
\frac{\mathrm{d}}{\mathrm{d}x} Q_n(x) = \frac{n+1}{x^2-1}\left(-x\,Q_n(x) + Q_{n+1}(x)\right) .
\end{equation}
Conversely, $E_{\perp,\mathrm{inside}}=0$, because the inside potential does not depend on $\eta$, see (\ref{eq:6}). Therefore the charge density $\sigma_q$ on the surface of the ellipsoid is equal to $\epsilon_0 E_{\perp,\mathrm{outside}}$ (Ref.~\onlinecite{Gauss-theorem}). 


The total charge $\mathrm{d}q_\mathrm{ell}$ on the surface of the ellipsoid in a ring of width $\mathrm{d}\theta$ at position $\theta$ is 
\begin{equation}\label{eq:47}\begin{split}
\mathrm{d}q_\mathrm{ell} = & \underbrace{\epsilon_0 E_{\perp,\mathrm{outside}}}_{=\sigma_q} \underbrace{2\pi a\sinh(\eta_0)\sin(\theta)}_{=p(\theta)} \underbrace{a\sqrt{(\sinh(\eta_0))^2+(\sin(\theta))^2} |\mathrm{d}\theta|}_{=\mathrm{d}w} \\
= 
& \frac{\epsilon_0 I_0 a\sin(\theta)|\mathrm{d}\theta|}{\kappa (c-a)}\sum_{n=0}^\infty \frac{4n+3}{2n+1} P_{2n+1}(\cos(\theta)) \,\left(\cosh(\eta_0) - \frac{Q_{2n+2}(\cosh(\eta_0))}{Q_{2n+1}(\cosh(\eta_0))}\right) ,
\end{split}\end{equation}
where $p(\theta)$ is the perimeter of the ring at position $\theta$, and $\mathrm{d}w$ is the width of the ring in the direction tangentially to the surface of the ellipse. If the slimness of the ellipsoid gets infinite, it means $\eta_0\to 0$ and $\mathrm{d}z\to -a\sin(\theta)\mathrm{d}\theta$ and $\cosh(\eta_0)=c/a\to 1$ and $\cos(\theta)=z/c$. In Appendix \ref{sec:appB} we prove 
\begin{equation}\label{eq:49m}\begin{split}
\lim_{\eta_0\to 0} \left(\cosh(\eta_0) - \frac{Q_{2n+2}(\cosh(\eta_0))}{Q_{2n+1}(\cosh(\eta_0))}\right) = \frac{-1}{(n+1)\ln\left(\cosh(\eta_0)-1\right)} .
\end{split}\end{equation}
With $\lambda_\mathrm{ell}=\mathrm{d}q_\mathrm{ell}/\mathrm{d}z$, we get for the line charge density of an infinitely slim ellipsoid 
\begin{equation}\label{eq:48}\begin{split}
\lim_{b\to 0} \lambda_\mathrm{ell} & = \frac{-\epsilon_0 I_0 }{\kappa (c-a)} \frac{1}{\ln(c/a-1)} \sum_{n=0}^\infty \frac{4n+3}{(n+1)(2n+1)} P_{2n+1}\!\left(\frac{z}{c}\right) .
\end{split}\end{equation}
The infinite sum in (\ref{eq:48}) can be summed up, see Appendix \ref{sec:appC}. With $c-a\to b^2/(2c)$ and $\ln(c/a-1)\to \ln(2) - 2\ln(2c/b)$, both valid for $b/c\to\infty$, we get 
\begin{equation}\label{eq:50}\begin{split}
\lim_{b\to 0} \lambda_\mathrm{ell}  = q_\mathrm{cont} \,\frac{c}{b^2} \frac{1}{\ln(2c/b)-\ln(2)/2} \,\ln\!\left(\frac{c+z}{c-z}\right) .
\end{split}\end{equation}
For $|z|<<c$ the line charge density is linear in $z$,
\begin{equation}\label{eq:51}\begin{split}
\lim_{b\to 0} \lambda_\mathrm{ell} & = q_\mathrm{cont} \, \frac{2}{b^2} \frac{z}{\ln(2c/b) - \ln(2)/2} + \mathcal{O}\left(z\right)^3 ,
\end{split}\end{equation}
which is identical to Ref.~\onlinecite{Assis1999b} for $b<<c$. 
If the wire becomes infinitely long, yet its diameter remains finite, the line charge density in (\ref{eq:51}) goes logarithmically to zero, which is also identical to Refs.~\onlinecite{Assis1999} and \onlinecite{CombesLaue1980}. If the wire length is fixed and the diameter goes to zero, the line charge density grows unboundedly. 
The total charge in the upper half of the infinitely thin wire is 
\begin{equation}\label{eq:55}\begin{split}
\lim_{b\to 0} q_\mathrm{ell} & = \int_0^c \lim_{b\to 0} \lambda_\mathrm{ell}\,\mathrm{d}z = \left(\frac{2c}{b}\right)^2 \frac{\ln(2)}{2\ln(2c/b)-\ln(2)} \,q_\mathrm{cont} > 0 .
\end{split}\end{equation}
Therefore \emph{in the case of an infinitely thin resistive wire carrying DC-current, the net charge becomes infinite} on the surface of the upper half of the wire. The reason for this singular behavior is that the voltage drop along the wire rises faster than the capacitance of the wire diminishes while the ellipsoid gets thinner. Note that the logarithimc singularity of $\lambda_\mathrm{ell}$ in $z=\pm c$ is \emph{not} responsible for the net charge becoming infinite: If we use (\ref{eq:51}) instead of (\ref{eq:50}) in the integration 
of (\ref{eq:55}), the net charge is only $2\ln2\approx 1.39$ times smaller. In particular the charge on the surface of the ellipsoid is much larger than the charge on the contact, $q_\mathrm{ell}>>q_\mathrm{cont}$. {This is important, because it justifies the neglection of $q_\mathrm{cont}$ when we compute the electric force between two wires in the next section.}

\section{The Electric Force between Two Infinitely Thin Resistive Wires with DC Currents}
\label{sec:force}

\begin{figure}[t]
\centering
\includegraphics[width=0.60\textwidth]{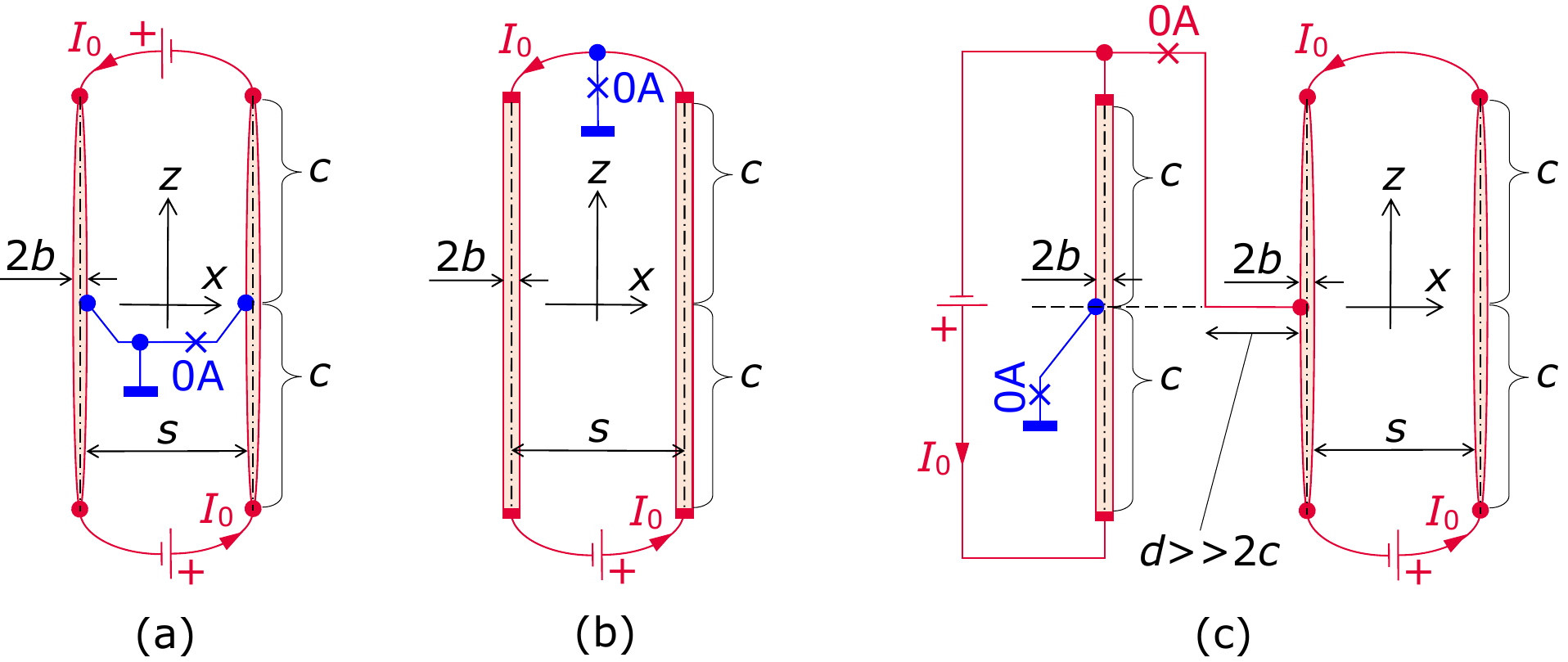}
\caption{Two identical thin parallel wires, spaced apart by a distance $s$, carry anti-parallel DC-currents $I_0$. The wires have lengths $2c$ and maximum diameters $2b$ in the limit $b\to 0$. The batteries are ideal current sources.
In scenario (a) the elliptical wires are grounded in the halves of the wires, i.e., in $z=0$. 
In scenario (b) the upper ends of the cylindrical wires are shorted and grounded in $z=c$. 
In the thin wire limit scenario (c) defines the same potential at the upper ends of the elliptical wires as scenario (b) does for cylindrical wires. 
}
\label{fig:two-parallel-wires1}
\end{figure}

{Let us consider two thin wires of finite lengths $2c$, both being parallel to the $z$-axis and extending from $z=-c$ to $z=c$. Their cross-sections are circular, and their center lines are spaced apart by a distance $s$. A DC-current $I_0$ flows in opposite directions through both wires. The wires are thought to be prolate ellipsoids with their thickest diameters in $z=0$ being $2b$. Let the wire diameter shrink infinitely, $\eta_0\to 0$, which means $b\to 0$ and $c\to a$. The asymptotic limit of this process is \emph{identical} to the thin wire limit of a \emph{cylindrical} wire of constant diameter $2b$, which also tends to zero, $2b\to 0$. During this limit process the line contacts of the ellipsoids shrink to point contacts as $c\to a$, whereas the circular disk contacts at the end faces of the cylindrical wires also become point contacts as $b\to 0$. \emph{For a finite diameter} $2b$ the resistance $R_\mathrm{ell}$ between the contacts of an elliptical wire is infinite ($\phi_\mathrm{inside}(\theta=0)=\infty$, see} (\ref{eq:6})), {whereas the resistance of a cylindrical wire is finite, $R_\mathrm{cyl}= 2c/(\pi\kappa b^2)$. However, \emph{in the thin wire limit} both resistances are identical, $R_\mathrm{cyl}\to R_\mathrm{ell}$, which means that the thinner the wires get the more similar their resistances become, although they both tend to infinity. Then the potentials along the wires, the electric field around them, the charges on them, and the forces between them converge to the very same limit. In the following we will see how the voltage between both wires and the choice of the common ground node affects the electric force.} 

\subsection{Wires are grounded at their halves: Figure \ref{fig:two-parallel-wires1}(a) }
\label{sec:scenario-a}

\begin{figure}[t]
\centering
\includegraphics[width=0.4\textwidth]{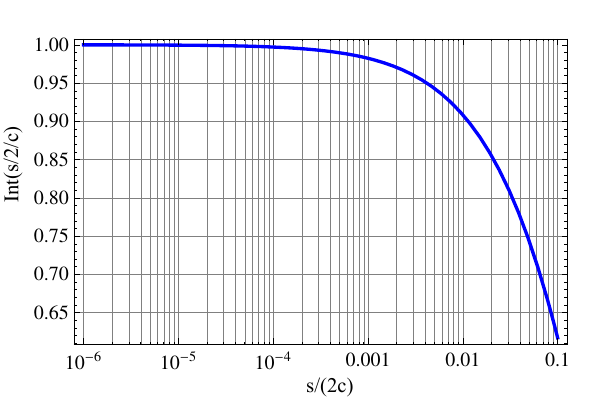}
\caption{Plot of the function $\mathrm{Int}(s/2/c)$ defined in (\ref{eq:69}). }
\label{fig:plot-of-Int}
\end{figure}

In this scenario a first current source is connected between the upper contacts of the wires, and a second identical current source is connected between their lower contacts (see Figure \ref{fig:two-parallel-wires1}(a)). Both wires are grounded in $z=0$. No current flows into the ground node due to the symmetry. 
Thus, the left wire has a line charge density $\lambda_\mathrm{ell}(z)$ from (\ref{eq:50}) and the right wire has a line charge density $(-1)\times\lambda_\mathrm{ell}(z)$. Then the $x$-component of the electric force on the line charge of the right wire is given by Coulomb's force law \cite{Coulomblaw} between two differential charges on the first and second wires, summed up over both wires, 
\begin{equation}\label{eq:60}\begin{split}
\lim_{b\to 0} F_{x,\mathrm{el}} & = \int_{z_1=-c}^c \int_{z_2=-c}^c \frac{-\lambda_\mathrm{ell}(z_1) \lambda_\mathrm{ell}(z_2)}{4\pi\epsilon_0} \,\frac{1}{s^2+(z_1-z_2)^2} \,\frac{s}{\sqrt{s^2+(z_1-z_2)^2}} \;\mathrm{d}z_1\,\mathrm{d}z_2 \\
& = \frac{-\epsilon_0 I_0^2 c^2 s}{\pi\kappa^2 b^4 \left(2\ln(2c/b)-\ln(2)\right)^2} \int_{-c}^c \int_{-c}^c  \frac{\ln\!\left(\frac{c+z_1}{c-z_1}\right) \ln\!\left(\frac{c+z_2}{c-z_2}\right)}{\left(s^2+(z_1-z_2)^2\right)^{3/2}}\,\mathrm{d}z_1\mathrm{d}z_2 .
\end{split}\end{equation}
This equation can be massaged into the following form (see Appendix \ref{sec:appE}),
\begin{equation}\label{eq:69}\begin{split}
& \lim_{b\to 0} F_{x,\mathrm{el}} = \frac{-2\pi\epsilon_0}{3}\,\frac{2c}{s}\,\left(\frac{I_0 c}{\kappa b^2 \left(2\ln(2c/b)-\ln(2)\right)}\right)^2 \,\mathrm{Int}\!\left(\frac{s}{2c}\right) < 0 , \\
& \mathrm{Int}\!\left(\frac{s}{2c}\right) = \frac{3}{\pi^2}\left(\frac{s}{2c}\right)^2 \int_{\alpha'=0}^1 \int_{\beta'=0}^{1-\alpha'} \frac{\ln\!\left(\frac{1+\alpha'+\beta'}{1-\alpha'-\beta'}\right) \ln\!\left(\frac{1-\alpha'+\beta'}{1+\alpha'-\beta'}\right)}{\left((\alpha')^2+(s/2/c)^2\right)^{3/2}} \,\mathrm{d}\alpha' \mathrm{d}\beta' .
\end{split}\end{equation}
The negative sign of $F_{x,\mathrm{el}}$ in (\ref{eq:69}) means that the electric force due to anti-parallel currents in both wires is attractive. Figure \ref{fig:plot-of-Int} shows a plot of $\mathrm{Int}(s/2/c)$. It is close to 1 if the wire spacing is less than 1\% of the wire length. 
Then, the electrical force between both wires is dominated by the factors in front of $\mathrm{Int}(s/2/c)$. 
In particular, for small spacing the electrical force is proportional to $c/s$, whereas for very large spacing, $s>>2c$, the electric force is proportional to $c^2/s^2$. For constant current the electric force grows unboundedly if the wire diameter diminishes. 


For anti-parallel currents $I_0$ in both wires, the magnetic force on the right wire is given by the Lorentz law, $
F_{x,\mathrm{magn}} = \mu_0 I_0^2 c /(\pi s) > 0 ,
$
with $\mu_0=4\pi\times 10^{-7}$ Vs/Am. The positive sign 
means 
repulsion. Finally, the ratio of electric over magnetic force is 
\begin{equation}\label{eq:77}
\lim_{b\to 0} \frac{F_{x,\mathrm{el}}}{F_{x,\mathrm{magn}}} = -\left(\frac{c L_\kappa}{b^2}\,\frac{1}{\ln(2c/b)-\ln(2)/2}\right)^2  \,\mathrm{Int}\!\left(\frac{s}{2c}\right) \quad\textrm{with } L_\kappa = \frac{\pi}{\sqrt{3}\kappa Z_0} ,
\end{equation}
wherein $Z_0 = \sqrt{\mu_0/\epsilon_0} \approx 376.7\;\Omega$ is the impedance of free space. The characteristic length $L_\kappa$ is $\approx\!1.8$ times the ratio of the wire resistivity over $Z_0$. For metal wires the resistivity in $\Omega\times\mathrm{m}$ is much smaller than $Z_0$ in $\Omega$, and therefore $L_\kappa$ is very small (fractions of a nano-meter, for copper wires $L_\kappa \approx 0.08\,\textrm{nm}$).  Consequently, the electric force is much smaller than the magnetic force, as long as the wire length is less than a few meters and the wire diameter is more than a tenth of a milli-meter (Table \ref{Tabelle1}). 

\begin{table}[h!]
\centering
\caption{Electric and magnetic forces between the wires in Figures \ref{fig:two-parallel-wires1}(a,b) in the thin wire limit. The wires are made of copper with $\kappa = 5.88\times 10^{7}\;\mathrm{S/m}$. The current density is always $100\,\mathrm{A}/\mathrm{mm}^2$, which gives the current $I_0$. The electrostatic induction between both wires is neglected (see Section \ref{sec:interference}). }
\begin{ruledtabular}
\begin{tabular}{c c c c c c c c}
\# & $c$ & $b$ & $s$ & $F_{x,\mathrm{magn}}$ for & $I_0$ & $F_{x,\mathrm{el}}/F_{x,\mathrm{magn}}$ & $F_{x,\mathrm{el}}/F_{x,\mathrm{magn}}$  \\
& & & & $100\,\mathrm{A}/\mathrm{mm}^2$ & & with (\ref{eq:77}), Fig. \ref{fig:two-parallel-wires1}(a) & with (\ref{eq:superpos7}), Fig. \ref{fig:two-parallel-wires1}(b) \\ 
\hline	
1 & 10 m & 0.1 mm & 1 cm & 3.95 mN & 3.14 A & -47.2 ppm & -99.2 ppm \\
2 & 1 m & 0.1 mm & 1 cm & 395 $\mu$N & 3.14 A & -0.692 ppm & -1.48 ppm \\
3 & 1 m & 0.1 mm & 10 cm & 39.5 $\mu$N & 3.14 A & -0.546 ppm & -1.34 ppm \\
4 & 1 m & 1 mm & 10 cm & 0.395 N & 314 A & $-9.5\times 10^{-11}$ & $-2.3\times 10^{-10}$ \\ 
5 & 0.5 m & 0.5 mm & 6 mm & 0.206 N & 78.5 A & $-4.8\times 10^{-10}$ & $-1.0\times 10^{-9}$ \\ 
6 & 1 m & 1 $\mu$m & 1 mm & $4\times 10^{-11}$ N & 314 $\mu$A & -33.08 & -70.24 \\
7 & 1 m & 1 nm & 1 mm & $4\times 10^{-23}$ N &  314 pA & $-1.5\times 10^{13}$ & $-3.2\times 10^{13}$ \\
\end{tabular}
\end{ruledtabular}
\label{Tabelle1}
\end{table}

\subsection{Wires are grounded at their upper ends: Figure \ref{fig:two-parallel-wires1}(b) }
\label{sec:superposition}

Here we compute the electric force between the two cylindrical wires of Figure \ref{fig:two-parallel-wires1}(b) in the limit $b\to 0$. The striking difference to the preceding Section is that now the wires are shorted at their upper ends.
In the thin wire limit the scenario in Figure \ref{fig:two-parallel-wires1}(b) is equivalent to the scenario in Figure \ref{fig:two-parallel-wires1}(c), which we use to compute the electric force. There the potential in the middle of the left elliptical wire is $\phi(x=-s/2,z=0)=-R_\mathrm{cyl} I_0/2$ and in the thin wire limit the potential in the middle of the right elliptical wire is $\phi(x=s/2,z=0)=R_\mathrm{cyl} I_0/2$ (because $R_\mathrm{cyl}\to R_\mathrm{ell}$). Thus, we may consider the potential in Scenario (c) as a linear superposition of the scenario in Figure \ref{fig:two-parallel-wires1}(a) and an electrostatic scenario.  In the electrostatic scenario no current flows through the ellipsoids, and they are charged up to $\pm R_\mathrm{cyl} I_0/2$. For the electric potentials it holds $\phi_{(c)}=\phi_{(a)}+\phi_\mathrm{static}$, where the indices '$(a)$, '$(c)$' refer to the Figures \ref{fig:two-parallel-wires1}(a),(c) and the index '$\mathrm{static}$' denotes the electrostatic case. The charges also add up analogously. The amount of charge needed to hold a prolate ellipsoid at potential $R_\mathrm{cyl} I_0/2$ is given  by
\begin{equation}\label{eq:superpos3}
q_\mathrm{static} = C_\mathrm{ell} R_\mathrm{cyl} \frac{I_0}{2} \quad\textrm{with } \lim_{b\to 0} C_\mathrm{ell} 
= \frac{4\pi\epsilon_0 c}{\ln\!\left(\frac{2c}{b}\right)} , 
\end{equation}
whereby the capacitance of a prolate ellipsoid, $C_\mathrm{ell}$, is derived in Appendix \ref{sec:app-capacitance}.  In the limit of an infinitely thin cylindrical wire it is known that the charge in electrostatic equilibrium ($I_0=0$) distributes uniformly on it \cite{Maxwell,Andrews,Jackson-revisited,Jackson2}. Hence, the line charge density in the electrostatic case is $\lambda_\mathrm{static} = q_\mathrm{static}/(2c)$ with 
\begin{equation}\label{eq:superpos5}
\lim_{b\to 0} \lambda_\mathrm{static} = \lim_{b\to 0} \pi \epsilon_0 \,\frac{I_0 R_\mathrm{cyl}}{\ln(2c/b)} .
\end{equation}
If we add $\lambda_\mathrm{static}$ to $\lambda_\mathrm{ell}$ in (\ref{eq:50}), we finally get the electric force between infinitely thin \emph{cylindrical} wires in Figure \ref{fig:two-parallel-wires1}(b)
\begin{equation}\label{eq:superpos7}
\lim_{b\to 0} F_{x,\mathrm{el}}  = -\pi\epsilon_0\,\frac{c}{s}\,\left(\frac{I_0 R_\mathrm{cyl}}{\ln(2c/b)}\right)^2 \,\left( \frac{\pi^2}{12} \,\mathrm{Int}\!\left(\frac{s}{2c}\right) + 1 \right) . 
\end{equation}
Comparison of (\ref{eq:superpos7}) with (\ref{eq:69}) shows that the electric force between infinitely thin wires grounded at their upper ends in Figure \ref{fig:two-parallel-wires1}(b) is 2.2 times stronger than if the wires are grounded at their centers in Figure \ref{fig:two-parallel-wires1}(a).

\section{{Electrostatic Induction} between Both Wires}
\label{sec:interference}

If the current carrying wires of length $2c$ are brought at a distance $s<<2c$, the surface charges on the first wire generate an electric field that acts on the second wire, and \emph{redistributes} the charges there. So far, we have neglected this electrostatic induction, but here we want to estimate its order of magnitude. {To this end we compare the electric field generated by the charges of a wire with the electric field generated by the charges of the other wire.} Thereby we only need to consider $E_z$, because in the thin wire limit the charges cannot move in lateral x-, y-directions.
\newline \underline{$E_z$ generated by the charges of the first wire onto themselves:} 
\newline In the thin wire limit we use $\eta=\mathrm{d}\eta$ with $|\mathrm{d}\eta|<<1$ in (\ref{eq:1}). It gives $\mathrm{d}r = a \sin(\theta) \mathrm{d}\eta$ and $z  = a \cos(\theta)$ for $|z|<a$. 
Inserting this into (\ref{eq:6}) and differentiating against $z$ gives 
\begin{equation}\label{eq:21}\begin{split}
E_z(r=0, |z|<a) & =  \frac{-I_0 a}{2\pi\kappa (c-a)}\;\frac{1}{a^2-z^2} 
\end{split}\end{equation}
In the limit of an infinitely thin wire we replace $a\to c$ and $c-a\to b^2/(2c)$ and get 
\begin{equation}\label{eq:90}\begin{split}
E_z(r=0, |z|<c) \to  \frac{-I_0 c^2}{\pi\kappa b^2}\;\frac{1}{c^2-z^2} .
\end{split}\end{equation}
\newline \underline{$E_z$ generated by the charges of one wire onto the other wire for the scenario in Figure \ref{fig:two-parallel-wires1}(a):}
\begin{equation}\label{eq:92}\begin{split}
E_z(r=s, z) & = \int_{z'=-c}^c \frac{\lambda_\mathrm{ell}(z')}{4\pi\epsilon_0} \,\frac{1}{s^2+(z-z')^2} \,\frac{z-z'}{\sqrt{s^2+(z-z')^2}} \;\mathrm{d}z' \\
& = \frac{I_0 c}{2\pi\kappa b^2 \left(2\ln(2c/b)-\ln(2)\right)}\;\frac{1}{\sqrt{(c^2+s^2-z^2)^2+(2sz)^2}} \\ 
& \quad\times \left\{ \sqrt{s^2+(c+z)^2}\ln\!\left(\frac{s^2+z^2-c^2+\sqrt{(c^2+s^2-z^2)^2+(2sz)^2}}{2\left(s^2+(c-z)^2\right)}\right) \right. \\
& \quad\quad \left. + \sqrt{s^2+(c-z)^2}\ln\!\left(\frac{s^2+z^2-c^2+\sqrt{(c^2+s^2-z^2)^2+(2sz)^2}}{2\left(s^2+(c+z)^2\right)}\right) \right\} .
\end{split}\end{equation}
This integral can be computed with Mathematica, but its exact formula is not even necessary if we pull out the diverging terms in $\lambda_\mathrm{ell}(z')$ for $b\to 0$, 
\begin{equation}\label{eq:94}\begin{split}
\lim_{b\to 0} \frac{E_z(r=0, |z|<c)}{E_z(r=s, z)} \propto \ln\!\left(\frac{2c}{b}\right) \to \infty .
\end{split}\end{equation}
This proves that the electric field acting on the charges of one wire produced by the charges on the other wire is infinitely smaller than the field of the charges on a wire on themselves, if both wires are infinitely thin while their spacing is finite. Therefore we can neglect {electrostatic induction} between both wires.

\section{Conclusion}

In this paper I discussed the surface charges on an infinitely thin straight resistive wire which carries a DC-current. Thereby the wire is replaced by a prolate ellipsoid in the limit of infinite slimness. The distribution of surface charges varies linearly with the position near the center of the wire, whereas it has a logarithmic singularity at the ends of the wire
. If two wires run parallel, 
electrostatic induction is negligible, as long as the 
wires are infinitely thin and spaced apart at a finite distance. The electric force between the surface charges on both wires was computed. It depends on the voltage drop along the wires, on the voltage between both wires, on the wire lengths, diameters, and their spacing. If the current, the wires lengths, and their spacing are fixed while the wires diameters shrink, the voltage drop grows unboundedly, and this will give infinite surface charges and infinite electric force. This limit leads to infinite current density and heating in the wire, which eventually distroys the wire. However, if we consider the \emph{ratio} of electric over magnetic force between both wires, this ratio is independent of the current, and we may scale down the current synchronously with the wire diameter to achieve constant current density during the limit process. Of course, the magnetic force decreases with the current accordingly, but in practice this just calls for a sufficiently sensitive method of force measurement. \emph{In such a scenario the electric force can indeed become even stronger than the magnetic force.} 

The electric force can be eliminated if each wire is surrounded by an electric shield (like a coaxial cable) and both shields are tied to the same potential (e.g. ground). Each shield has to be clamped mechanically to its conductor, because there might be an electric force between the shield and its conductor (depending on symmetry).

This manuscript was submitted to the American Journal of Physics, but it was rejected (too long, too mathematically dense, lack of interest).

The author has no conflicts to disclose.

\appendix

\section{How to determine the constant $c_1$ of the potential $\phi_\mathrm{inside}$}
\label{sec:appL}

Let us look at the potential close to $z=0$, which means $z=0+\mathrm{d}z$ with $|\mathrm{d}z|<<a$ and $\theta=\pi/2+\mathrm{d}\theta$ with $|\mathrm{d}\theta|<<1$. Inserting this into (\ref{eq:1}) gives 
\begin{equation}\label{eq:10}\begin{split}
x & \approx a \sinh(\eta) \cos(\psi) , \quad 
y \approx= a \sinh(\eta) \sin(\psi) , \quad 
\mathrm{d}z  \approx -a \cosh(\eta) \,\mathrm{d}\theta . 
\end{split}\end{equation}
With the radial distance $r=\sqrt{x^2+y^2}$ it follows with (\ref{eq:10}) 
\begin{equation}\label{eq:11}
\left(\cosh(\eta)\right)^2 = 1 + \left(\sinh(\eta)\right)^2 \quad\Rightarrow\quad 
\left(\frac{\mathrm{d}z}{a \,\mathrm{d}\theta}\right)^2 = 1 + \left(\frac{r}{a}\right)^2 .
\end{equation}
We solve (\ref{eq:11}) for $\mathrm{d}\theta$, whereby $\mathrm{d}\theta$ and $\mathrm{d}z$ have opposite sign. Inserting this into (\ref{eq:6}) gives 
\begin{equation}\label{eq:12}
\mathrm{d}\phi_\mathrm{inside}(r,\mathrm{d}z) = \frac{c_1}{2} \ln\!\left(\frac{1-\mathrm{d}\theta}{1+\mathrm{d}\theta}\right) = \frac{c_1}{2} \ln\!\left(\frac{1+\mathrm{d}z/\sqrt{a^2+r^2}}{1-\mathrm{d}z/\sqrt{a^2+r^2}}\right) = \frac{c_1\,\mathrm{d}z}{\sqrt{a^2+r^2}} .
\end{equation}
From (\ref{eq:12}) we get the $z$-component of the electric field in $z=0$ 
\begin{equation}\label{eq:13}
E_z(r,z=0) = -\left.\frac{\partial\phi_\mathrm{inside}(r,z)}{\partial z}\right|_{z=0} = \frac{-c_1}{\sqrt{a^2+r^2}} < 0 .
\end{equation}
With Ohm's law the $z$-component of the current density is $J_z = \kappa E_z$, and the total current $I_0$ downward through the ellipsoid is given by the integral 
\begin{equation}\label{eq:15}\begin{split}
& I_0 = \int_{r=0}^{b} (-J_z) 2\pi r\,\mathrm{d}r = 2\pi\kappa c_1  \int_{r=0}^{b} \frac{r}{\sqrt{a^2+r^2}}\,\mathrm{d}r = 2\pi\kappa c_1 (c-a) , \\
& \Rightarrow \lim_{b\to 0} c_1 = \lim_{b\to 0} \frac{I_0}{2\pi\kappa (c-a)} = \frac{I_0 \,c}{\pi\kappa\,b^2} .
\end{split}\end{equation}
Here we used $a^2+b^2=c^2$. Inserting $c_1$ from (\ref{eq:15}) into (\ref{eq:6}) gives the potential everywhere inside the ellipsoid, if the current is known. 

\section{Proof of (\ref{eq:42b})}
\label{sec:appA}

\begin{equation}\label{eq:appA1}\begin{split}
& \int_{-1}^1 P_{2n+1}(x) Q_0(x) \,\mathrm{d}x = \int_{-1}^1 P_{2n+1}(x) \frac{1}{2}\ln\!\left(\frac{1+x}{1-x}\right) \mathrm{d}x \\ 
& = \int_{-1}^1 P_{2n+1}(x) \sum_{m=0}^\infty \frac{x^{2m+1}}{2m+1} \,\mathrm{d}x \\ 
& = \sum_{m=0}^\infty  \int_{-1}^1 \frac{P_{2n+1}(x)}{2m+1}\sum_{\ell=0}^m \frac{2^{2\ell +1} (4\ell+3) (2m+1)! (m+\ell+1)!}{(2m+2\ell+3)! (m-\ell)!}P_{2\ell +1}(x) \,\mathrm{d}x ,
\end{split}\end{equation}
where we first developped the logarithm into a Taylor series, and then we used the Legendre series/sum of odd powers of $x$ from Ref.~\onlinecite{Arfken2}. Next we apply the orthogonality of the Legendre polynomials on (\ref{eq:appA1}), $\int_{-1}^1 P_{2n+1}(x) P_{2\ell +1}(x) \,\mathrm{d}x = 2\delta_{\ell ,n}/(4n+3)$, and we reverse the sequence of summations, $ \sum_{m=0}^\infty \sum_{\ell=0}^m = \sum_{\ell=0}^\infty \sum_{m=\ell}^\infty$. Then (\ref{eq:appA1}) becomes 
\begin{equation}\label{eq:appA3}\begin{split}
& = 2^{2n+2} \sum_{m=n}^\infty \frac{(2m)! (m+n+1)!}{(2m+2n+3)! (m-n)!} = \frac{1}{(n+1)(2n+1)} := R_n .
\end{split}\end{equation}
The summation in (\ref{eq:appA3}) is handled by Mathematica. 
For its proof we start with the following identity \cite{Arfken5}
\begin{equation}\label{eq:appA5}
Q_{2n+1}(x) = 2^{2n+1}\sum_{\ell=0}^\infty \frac{(2\ell+2n+1)!\,(\ell+2n+1)!}{(2\ell+4n+3)!\, \ell!} x^{-2\ell-2n-2} .
\end{equation}
Integration of (\ref{eq:appA5}) gives 
\begin{equation}\label{eq:appA6}\begin{split}
& \int_1^\infty Q_{2n+1}(x)\,\mathrm{d}x \\ 
& = 2^{2n+1}\sum_{\ell=0}^\infty \frac{(2\ell+2n)!\,(\ell+2n+1)!}{(2\ell+4n+3)!\,\ell!} \underbrace{(2\ell+2n+1) \int_1^\infty x^{-2\ell-2n-2}\,\mathrm{d}x}_{=1} = \frac{1}{2} R_n .
\end{split}\end{equation}
With the recurrence relation \cite{Arfken6}
\begin{equation}\label{eq:appA7}
(2n+1) Q_n(x) = Q'_{n+1}(x) - Q'_{n-1}(x) ,
\end{equation}
it follows for $0<\epsilon<<1$ 
\begin{equation}\label{eq:appA8}
\int_{1+\epsilon}^\infty Q_{2n+1}(x)\,\mathrm{d}x = \int_{1+\epsilon}^\infty \frac{Q'_{2n+2}(x)-Q'_{2n}(x)}{4n+3}\,\mathrm{d}x = \frac{Q_{2n}(1+\epsilon)-Q_{2n+2}(1+\epsilon)}{4n+3} .
\end{equation}
In (\ref{eq:appA7},\ref{eq:appA8}) the primes denote differentiation with respect to $x$. We insert (\ref{eq:appB5}) into (\ref{eq:appA8}), let $\epsilon\to 0$, and insert this into (\ref{eq:appA6}), 
\begin{equation}\label{eq:appA9}
\frac{1}{2} R_n = \frac{\omega_{2n+2}-\omega_{2n}}{4n+3} .
\end{equation}
Inserting (\ref{eq:appB14}) and (\ref{eq:appB18}) into (\ref{eq:appA9}) finally gives 
\begin{equation}\label{eq:appA10}\begin{split}
R_n & = 2 \frac{(\omega_{2n+2}-\omega_{2n+1})+(\omega_{2n+1}-\omega_{2n})}{4n+3} \\ 
& = \frac{2}{4n+3}\left( \frac{1}{2n+2} + \frac{1}{2n+1} \right) =\frac{1}{(n+1)(2n+1)} .
\end{split}\end{equation}

\section{Proof of (\ref{eq:49m})}
\label{sec:appB}

With Mathematica we compute 
\begin{equation}\label{eq:appB1}
\lim_{x\to 1+0} \left(x - \frac{Q_{2n+2}(x)}{Q_{2n+1}(x)}\right) \ln(x-1) = \frac{-1}{n+1} .
\end{equation}
For a proof we start with the recurrence relation \cite{Arfken6,Hobson1} 
\begin{equation}\label{eq:appB3}\begin{split}
& n Q_n(x) = (2n-1) x Q_{n-1}(x) - (n-1) Q_{n-2}(x) , \\
& \Rightarrow \lim_{x\to 1+0} n Q_n(x) = \lim_{x\to 1+0}  (2n-1) Q_{n-1}(x) + (n-1) Q_{n-2}(x)
\end{split}\end{equation}
with the limits of the first three Q-functions 
\begin{equation}\label{eq:appB4}\begin{split}
& \lim_{x\to 1+0} Q_0(x) = \lim_{x\to 1+0} \frac{1}{2}\ln\!\left(\frac{2}{x-1}\right) , \\ 
& \lim_{x\to 1+0} Q_1(x) = \lim_{x\to 1+0} \frac{1}{2}\ln\!\left(\frac{2}{x-1}\right) - 1 , \\ 
& \lim_{x\to 1+0} Q_2(x) = \lim_{x\to 1+0} \frac{1}{2}\ln\!\left(\frac{2}{x-1}\right) - \frac{3}{2} . 
\end{split}\end{equation}
Inserting (\ref{eq:appB4}) into (\ref{eq:appB3}) shows that the logarithmic term is identical for all $Q_n(x\to 1+0)$. Thus, we can write 
\begin{equation}\label{eq:appB5}\begin{split}
& \lim_{x\to 1+0} Q_n(x) = \lim_{x\to 1+0} \frac{1}{2}\ln\!\left(\frac{2}{x-1}\right) - \omega_n , 
\end{split}\end{equation}
where $\omega_n$ is a rational number. We insert (\ref{eq:appB5}) into the recurrence relation (\ref{eq:appB3}) and get 
\begin{equation}\label{eq:appB6}\begin{split}
& n \omega_n = (2n-1) \omega_{ n-1} - (n-1) \omega_{n-2} , \quad\textrm{with } \omega_0=0, \omega_1=1, \omega_2=3/2 . 
\end{split}\end{equation}
Inserting (\ref{eq:appB6}) into the left side of (\ref{eq:appB1}) gives 
\begin{equation}\label{eq:appB10}\begin{split}
& \lim_{x\to 1+0} \frac{Q_{2n+2}(x)}{Q_{2n+1}(x)} = \lim_{x\to 1+0} 1 + 2\frac{\omega_{2n+2} - \omega_{2n+1}}{\ln(x-1)} , \\ 
\Rightarrow & \lim_{x\to 1+0} \left(x - \frac{Q_{2n+2}(x)}{Q_{2n+1}(x)}\right) \ln(x-1) = -2\left(\omega_{2n+2} - \omega_{2n+1}\right) .
\end{split}\end{equation}
If we express $\omega_{2n+2}$ by $\omega_{2n+1}, \omega_{2n}$ via the recurrence relations (\ref{eq:appB6}) we get after a few re-arrangements 
\begin{equation}\label{eq:appB12}
(2n+2) (\omega_{2n+2} - \omega_{2n+1}) = 2n (\omega_{2n} - \omega_{2n-1}) .
\end{equation}
For the first few indices $n$ this gives 
\begin{equation}\label{eq:appB14}\begin{split}
n=1: \; & 4(\omega_4-\omega_3) = 2(\omega_2-\omega_1) = 2(\frac{3}{2}-1) = 1 , \\
n=2: \; & 6(\omega_6-\omega_5) = 4(\omega_4-\omega_3) = 1 , \\
n=3: \; & 8(\omega_8-\omega_7) = 6(\omega_6-\omega_5) = 1 , \\
n: \; & \omega_{2n+2}-\omega_{2n+1} = \frac{1}{2n+2} .
\end{split}\end{equation}
Inserting this into (\ref{eq:appB10}) completes the proof of (\ref{eq:appB1}).

Moreover, setting $n\to 2n+1$ in (\ref{eq:appB6}) gives 
\begin{equation}\label{eq:appB16}
(2n+1) (\omega_{2n+1} - \omega_{2n}) = 2n (\omega_{2n} - \omega_{2n-1}) .
\end{equation}
For the first few indices $n$ this gives 
\begin{equation}\label{eq:appB18}\begin{split}
n=1: \; & 3(\omega_3-\omega_2) = 2(\omega_2-\omega_1) = 2(\frac{3}{2}-1) = 1 , \\
n=2: \; & 5(\omega_5-\omega_4) = 4(\omega_4-\omega_3) = 1 , \\
n=3: \; & 7(\omega_7-\omega_6) = 6(\omega_6-\omega_5) = 1 , \\
n: \; & \omega_{2n+1}-\omega_{2n} = \frac{1}{2n+1} .
\end{split}\end{equation}

\section{How to compute the sum in (\ref{eq:48})}
\label{sec:appC}

We want to compute 
\begin{equation}\label{eq:appC0}
S = \sum_{n=0}^\infty \frac{4n+3}{2n+1} \,\frac{P_{2n+1}(x)}{n+1} = 2 \underbrace{\sum_{n=0}^\infty \frac{P_{2n+1}(x)}{n+1}}_{=S_1} + \underbrace{\sum_{n=0}^\infty \frac{1}{2n+1}\,\frac{P_{2n+1}(x)}{n+1}}_{=S_2} .
\end{equation}
For the sum $S_1$ we start with the generating function of the Legendre polynomials \cite{Arfken4}
\begin{equation}\label{eq:appC3}
\frac{1}{\sqrt{1-2xt+t^2}} = \sum_{n=0}^\infty P_n(x) t^n \quad\forall t\in [-1,1] .
\end{equation}
We integrate (\ref{eq:appC3}) once over $t:0\to 1$ and once over $t:0\to -1$. Adding both results cancels out even indices $n$,
\begin{equation}\label{eq:appC5}\begin{split}
& \int_0^1 \frac{1}{\sqrt{1-2xt+t^2}}\,\mathrm{d}t + \int_0^{-1} \frac{1}{\sqrt{1-2xt+t^2}}\,\mathrm{d}t \\ 
& \quad = \ln\!\left(1+\sqrt{\frac{2}{1-x}}\right) - \ln\!\left(1+\sqrt{\frac{2}{1+x}}\right) \\
& \quad = \sum_{\mathrm{odd}\;n} P_n(x) \frac{1+(-1)^{n+1}}{n+1}= \sum_{n=0}^\infty \frac{P_{2n+1}(x)}{n+1} = S_1 . 
\end{split}\end{equation}
For the sum $S_2$ we set $t=yz$ in (\ref{eq:appC3}) and divide both sides by $y$. This gives 
\begin{equation}\label{eq:appC7}
\frac{1}{y\sqrt{1-2xyz+y^2 z^2}} = \sum_{n=0}^\infty P_n(x) y^{n-1} z^n \quad\forall y,z\in [-1,1] .
\end{equation}
Like above, we integrate (\ref{eq:appC7}) once over $z:0\to 1$ and once over $z:0\to -1$, and we add both results to cancel out even indices $n$, 
\begin{equation}\label{eq:appC9}\begin{split}
& \int_0^1 \frac{1}{y\sqrt{1-2xyz+y^2 z^2}}\,\mathrm{d}z + \int_0^{-1} \frac{1}{y\sqrt{1-2xyz+y^2 z^2}}\,\mathrm{d}z \\ 
& \quad = \sum_{\mathrm{odd}\;n} P_n(x) y^{n-1} \frac{1+(-1)^{n+1}}{n+1}= \sum_{n=0}^\infty \frac{P_{2n+1}(x)}{n+1}\,y^{2n} . 
\end{split}\end{equation}
Next, we integrate (\ref{eq:appC9}) over $y:-1\to 1$ whereby we take the Cauchy principal value around the singularity in $y=0$, 
\begin{equation}\label{eq:appC10}\begin{split}
& \fint_{-1}^1\left( \int_0^1 \frac{1}{y\sqrt{1-2xyz+y^2 z^2}}\,\mathrm{d}z + \int_0^{-1} \frac{1}{y\sqrt{1-2xyz+y^2 z^2}}\,\mathrm{d}z \right) \mathrm{d}y \\ 
& \quad = 2\ln\!\left(\frac{3+x+2\sqrt{2}\sqrt{1+x}}{3-x+2\sqrt{2}\sqrt{1-x}}\right)  \\ 
& \quad = \sum_{n=0}^\infty \frac{P_{2n+1}(x)}{n+1}\,\frac{1-(-1)^{2n+1}}{2n+1} = 2 S_2 . 
\end{split}\end{equation}
The computation of the integrals in (\ref{eq:appC10}) is easier, if one starts with the integration over $y$ before the integration over $z$, 
\begin{equation}\label{eq:appC11}\begin{split}
& \fint_{-1}^1 \frac{1}{y\sqrt{1-2xyz+y^2 z^2}}\,\mathrm{d}y = \ln\!\left(\frac{1+xz+\sqrt{1+2xz+z^2}}{1-xz+\sqrt{1-2xz+z^2}}\right) , \\ 
& \int_0^1 \ln\!\left(\frac{1+xz+\sqrt{1+2xz+z^2}}{1-xz+\sqrt{1-2xz+z^2}}\right) \mathrm{d}z = \ln\!\left(\frac{3+x+2\sqrt{2}\sqrt{1+x}}{3-x+2\sqrt{2}\sqrt{1-x}}\right) .
\end{split}\end{equation}
Combining (\ref{eq:appC5}) and (\ref{eq:appC10}) finally gives 
$S = 2 S_1 + S_2 = \ln(1+x) - \ln(1-x)$. 

\section{How to derive $\mathrm{Int}\!\left(\frac{s}{2c}\right)$ in equation (\ref{eq:69})}
\label{sec:appE}

\begin{figure}[t]
\centering
\includegraphics[width=0.2\textwidth]{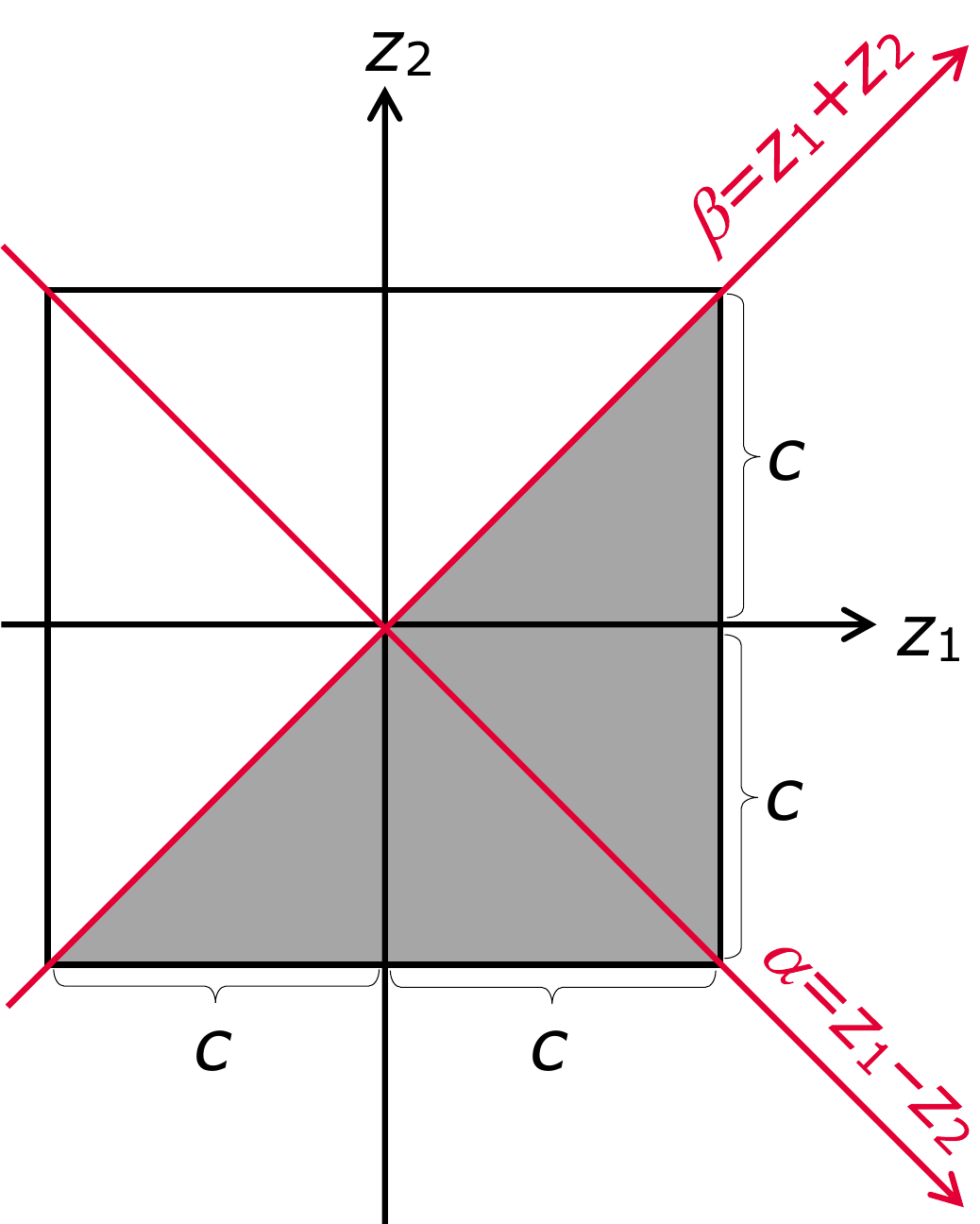}
\caption{Transformation of coordinates $(z_1,z_2)\mapsto (\alpha,\beta)$ to compute the integral in (\ref{eq:appE67}). }
\label{fig:coordinate-transformation1}
\end{figure}

Let us call the integrand in the last line of (\ref{eq:60}) $f(z_1,z_2)$. It is symmetrical, because it is identical to $f(z_2,z_1)$. Therefore, we may halve the integration domain in (\ref{eq:60}), 
\begin{equation}\label{eq:appE62}
\int_{z_1=-c}^c \int_{z_2=-c}^c f(z_1,z_2)\,\mathrm{d}z_1\mathrm{d}z_2 = 2 \int_{z_1=-c}^c \int_{z_2=-c}^{z_1} f(z_1,z_2)\,\mathrm{d}z_1\mathrm{d}z_2 .
\end{equation}
In Figure \ref{fig:coordinate-transformation1} this reduces the integration domain from the square $z_1\in [-c,c] \land z_2\in [-c,c]$ to the dark triangle. Next, we transform the integration variables $(z_1,z_2)\mapsto (\alpha,\beta)$ according to
\begin{equation}\label{eq:appE64}
\alpha = z_1-z_2, \; \beta = z_1+z_2, \quad\leftrightarrow\quad z_1=\frac{\alpha+\beta}{2}, \; z_2=\frac{-\alpha+\beta}{2}, 
\end{equation}
which is also shown in Figure \ref{fig:coordinate-transformation1}. Thereby the differential surface elements relate via the Jacobian determinant,  
\begin{equation}\label{eq:appE65}
\mathrm{d}z_1\mathrm{d}z_2 = \left|\frac{\partial(z_1,z_2)}{\partial(\alpha,\beta)}\right| \mathrm{d}\alpha\mathrm{d}\beta, \quad\textrm{with } \frac{\partial(z_1,z_2)}{\partial(\alpha,\beta)} = \left|
\begin{matrix}
	 1/2 & -1/2 \\
	1/2 & 1/2 \\
\end{matrix}
\right| = \frac{1}{2} .
\end{equation}
It holds 
\begin{equation}\label{eq:appE67}\begin{split}
& 2 \int_{z_1=-c}^c \int_{z_2=-c}^{z_1} f(z_1,z_2)\,\mathrm{d}z_1\mathrm{d}z_2 \\ 
& = 2 \int_{\alpha=0}^{2c} \left( \int_{\beta=\alpha -2c}^0 g(\alpha,\beta) \,\frac{\mathrm{d}\alpha\mathrm{d}\beta}{2} + \int_{\beta=0}^{2c-\alpha} g(\alpha,\beta) \,\frac{\mathrm{d}\alpha\mathrm{d}\beta}{2} \right) \\ 
& \textrm{with } g(\alpha,\beta) = \ln\!\left(\frac{2c+\alpha+\beta}{2c-\alpha-\beta}\right) \ln\!\left(\frac{2c-\alpha+\beta}{2c+\alpha-\beta}\right) \left(s^2+\alpha^2 \right)^{-3/2} .
\end{split}\end{equation}
$g(\alpha,\beta)$ is an even function of $\beta$. Therefore, in (\ref{eq:appE67}) the two integrals over $g(\alpha,\beta)$are identical, and the final result is given in (\ref{eq:69}), whereby we used the transformations $\alpha = 2c \alpha'$ and $\beta = 2c \beta'$. In (\ref{eq:69}) the integration over $\beta'$ can be done in closed form, 
\begin{equation}\label{eq:appE70}\begin{split}
\mathrm{Int}(x) = & \frac{x^2}{\pi^2} \int_{\alpha'=0}^1 
\frac{f_1(\alpha')}{\left({\alpha'}^2+x^2\right)^{3/2}} \,\mathrm{d}\alpha' , \\ 
f_1(\alpha') = & 2\pi^2 -3(1-\alpha')\left(\ln(1-\alpha')\right)^2-3\alpha'\left(\ln(\alpha')\right)^2+6\alpha'\mathrm{Li}_2(\alpha') \\ 
& -6(1+\alpha')\left(\mathrm{Li}_2(1+\alpha') + \imath\pi\ln(1+\alpha')\right) ,
\end{split}\end{equation}
with the imaginary unit $\imath=\sqrt{-1}$, and with the poly-logarithm $\mathrm{Li}_2(x)=\sum_{k=1}^\infty x^k/k^2$. $f_1(\alpha')$ is real, all imaginary portions are in the last line of (\ref{eq:appE70}) and cancel out. 

We compute the limit of $\mathrm{Int}(x)$ for $x\to 0$ with partial integration, 
\begin{equation}\label{eq:appE72}\begin{split}
& x^2 \int_{\alpha'=\epsilon}^{1-\epsilon} \frac{f_1(\alpha')}{\left({\alpha'}^2+x^2\right)^{3/2}} \,\mathrm{d}\alpha' = 
\left.\frac{\alpha' f_1(\alpha')}{\sqrt{\alpha'^2+x^2}}\right|_\epsilon^{1-\epsilon} - \int_{\alpha'=\epsilon}^{1-\epsilon} \frac{\alpha' f_2(\alpha')}{\sqrt{{\alpha'}^2+x^2}} \,\mathrm{d}\alpha' , \\ 
& f_2(\alpha') = \frac{\mathrm{d}f_1(\alpha')}{\mathrm{d}\alpha'} \\ 
& \qquad = 3\left( \left(\ln(1-\alpha')\right)^2-\left(\ln(\alpha')\right)^2-2\imath\pi\ln(1+\alpha')+2\mathrm{Li}_2(\alpha') - 2\mathrm{Li}_2(1+\alpha') \right) .
\end{split}\end{equation}
For the integral on the right hand side of in (\ref{eq:appE72}) we get again with partial integration 
\begin{equation}\label{eq:appE74}\begin{split}
\left.\sqrt{{\alpha'}^2+x^2} f_2(\alpha')\right|_\epsilon^{1-\epsilon} - \int_{\alpha'=\epsilon}^{1-\epsilon} \frac{6}{\alpha'}\sqrt{{\alpha'}^2+x^2}\,\frac{(1+\alpha')\ln(1-\alpha')+(1-\alpha')\ln(\alpha')}{{\alpha'}^2-1} \,\mathrm{d}\alpha' .
\end{split}\end{equation}
For $x\to 0$, the integral in (\ref{eq:appE74}) tends to $\pi^2/2+6\epsilon (\ln(\epsilon)-1) +3(\ln(\epsilon))^2$ (computed with Mathematica). We re-insert this into (\ref{eq:appE74}) and (\ref{eq:appE72}), and compute with Mathematica the limit for $\epsilon\to 0$. The result is 
$\lim_{x\to 0} \mathrm{Int}(x) = 1$.

\section{The Capacitance of a Prolate Ellipsoid}
\label{sec:app-capacitance}

The potential outside a charged metallic ellipsoid has the same ansatz as in (\ref{eq:40}) with the only non-zero coefficient 
$A_0 = \phi_0 / Q_0(\cosh(\eta_0))$ , 
whereby the ellipsoid is at potential $\phi_0$ and its surface has the ellipsoidal coordinate $\eta=\eta_0$. The electric field on the surface of the ellipsoid is given analogous to (\ref{eq:45}), 
\begin{equation}\label{eq:app-capacitance2}
E_{\perp,\mathrm{outside}} = \frac{-1}{a\sqrt{(\sinh(\eta_0))^2+(\sin(\theta))^2}}A_0 \underbrace{P_0(\cos(\theta))}_{=1} \underbrace{\frac{\mathrm{d}}{\mathrm{d}\eta_0} Q_0(\cosh(\eta_0))}_{=-1/\sinh(\eta_0)} .  
\end{equation}
The charge on this ellipsoid is given analogous to (\ref{eq:47}) 
\begin{equation}\label{eq:app-capacitance3}
q_\mathrm{ell} = \int_0^\pi \mathrm{d}q_\mathrm{ell}\,\mathrm{d}\theta = \frac{\epsilon_0 \phi_0}{Q_0(\cosh(\eta_0))} 2\pi a\int_0^\pi \sin(\theta)\,\mathrm{d}\theta .  
\end{equation}
We insert $a=\sqrt{c^2-b^2}$ into (\ref{eq:app-capacitance3}) and compute $q_\mathrm{ell}/\phi_0$, which gives $C_\mathrm{ell}$ from (\ref{eq:superpos3}), which is consistent to Refs. \onlinecite{Kottler,Smythe,Stratton2}. 





\begin{thebibliography}{99}
\bibitem{Assis1999b} see Ref.~\onlinecite{Assis1999}, eq.(12). 
\textit{Note that eq.(12) has a typo: the denominator should be multiplied by the wire length---compare with eq.(16) there.}
\end{thebibliography}
\end{document}